\documentclass[prb,twocolumn,aps,floatfix,showpacs,tightenlines,showpacs,superscriptaddress]{revtex4}
\usepackage{amssymb,amsmath,amsthm,amsbsy,epsfig,color,graphicx,times}

\newcommand{\beq}{\begin{eqnarray}}
\newcommand{\eeq}{\end{eqnarray}}
\newcommand{\be}{\begin{equation}}
\newcommand{\ee}{\end{equation}}
\newcommand{\ket}[1]{ \mid#1\rangle}

\newcommand{\Tr}{\mathrm{Tr}}

\begin{document}

\title{Universal aspects in the behavior of the entanglement spectrum in one dimension: \\
       scaling transition at the factorization point and ordered entangled structures}

\author{S. M. Giampaolo}
\affiliation{University of Vienna, Faculty of Physics, Boltzmanngasse 5, 1090 Vienna, Austria}

\author{S. Montangero}
\affiliation{Institut f\"{u}r Quanteninformationsverarbeitung,
Universit\"{a}t Ulm, D-89069 Ulm, Germany}

\author{F. Dell'Anno}
\affiliation{Liceo Statale P. E. Imbriani, via Pescatori 155, I-83100 Avellino, Italy}


\author{S. De Siena}
\affiliation{Dipartimento di Ingegneria Industriale, Universit\`a degli Studi di Salerno,
Via Giovanni Paolo II, I-84084 Fisciano (SA), Italy}
\affiliation{CNISM - Consorzio Nazionale Interuniversitario per le Scienze Fisiche della Materia, Unit\`a di Salerno, I-84084 Fisciano (SA), Italy}

\author{F. Illuminati}
\thanks{Corresponding author: illuminati@sa.infn.it}
\affiliation{Dipartimento di Ingegneria Industriale, Universit\`a degli Studi di Salerno,
Via Giovanni Paolo II, I-84084 Fisciano (SA), Italy}
\affiliation{CNISM - Consorzio Nazionale Interuniversitario per le Scienze Fisiche della Materia, Unit\`a di Salerno, I-84084 Fisciano (SA), Italy}

\date{September 17, 2013}


\begin{abstract}
We investigate the scaling of the entanglement spectrum and of the R\'enyi block entropies and determine its universal aspects in the ground state of critical and noncritical one-dimensional quantum spin models.
In all cases, the scaling exhibits an oscillatory behavior that terminates at the factorization point and whose frequency is universal. Parity effects in the scaling of the R\'enyi entropies for gapless models at zero field are thus shown to be a particular case of such universal behavior. Likewise, the absence of oscillations for the Ising chain in transverse field is due to the vanishing value of the factorizing field for this particular model. In general, the transition occurring at the factorizing field between two different scaling regimes of the entanglement spectrum corresponds to a quantum transition to the formation of finite-range, ordered structures of quasi-dimers, quasi-trimers, and quasi-polymers. This entanglement-driven transition is superimposed to and independent of the long-range magnetic order in the broken symmetry phase. Therefore, it conforms to recent generalizations that identify and classify the quantum phases of matter according to the structure of ground-state entanglement patterns. We characterize this form of quantum order by a global order parameter of entanglement defined as the integral, over blocks of all lengths, of the R\'enyi entropy of infinite order. Equivalently, it can be defined as the integral of the bipartite single-copy or geometric entanglement. The global entanglement order parameter remains always finite at fields below the factorization point and vanishes identically above it.
\end{abstract}

\pacs{75.10.Jm, 03.65.Ud, 03.67.Mn, 05.50.+q}

\maketitle

\section{Introduction}

During the past decade the use of entanglement in the study of complex quantum systems has developed at an increasingly fast
pace~\cite{AmicoFazioOsterlohVedral2008,EisertPlenioCramer2010,CalabreseCardyDoyon2009,Ladd2010,Osterloh2006}.
Extended analyses have established the monotonic scaling of the von Neumann entropy in the ground state of spin
models~\cite{VidalLatorreRicoKitaev2003,LatorreRicoVidal2004}, the so called {\em area-law}, and its profound relations both with conformal
field theory (CFT)~\cite{HolzheyLarsenWilczek1994,CalabreseCardy2004,CalabreseCardy2009} and  with the so-called ``majorization'' of the
entanglement along renormalization group flows in quantum spin chains~\cite{Latorre2005,Orus2005}.

The von Neumann entanglement entropy is in general extremely hard to compute and its measurement requires complete state
tomography~\cite{AmicoFazioOsterlohVedral2008}. On the other hand, it has been argued that the R\'enyi entropies of higher order contain
substantial information about the universal properties of a quantum many-body system~\cite{Zanardi2012,Hamma2012,Flammiaetal2009}. In particular,
the R\'enyi entropy of order $2$ is directly related to the purity of the ground-state block reduced density matrix, is significantly easier to
compute compared to the von Neumann entropy (i.e. the R\'enyi entropy of order $1$), and can in principle be measured
directly~\cite{Horodecki2002,HastingsGonzalezKallinMelko2010}.
Moreover, characterizing the behavior of the R\'enyi block entanglement entropies is equivalent to characterizing the entanglement spectrum,
i.e. the entire set of eigenvalues of the ground-state reduced block density matrix~\cite{Li2008,Peschel2009,Bernevig2010}.

Indeed, these quantities play a relevant role in determining the scaling properties of
numerical algorithms based on matrix product states~\cite{Shuchetal2008,Tagliacozzoetal2008,Pollmannetal2009,VerstraeteCirac2009}. The R\'enyi
entropies are also a useful tool to determine the continuous or discontinuous nature of a phase
transition~\cite{ErcolessiEvangelistiFranchiniRavanini2011} and to estimate quasi-long-range order in low-dimensional
systems~\cite{DalmonteErcolessiTaddia2011}. Furthermore, the concept of topological entanglement entropy~\cite{Kitaev2006,Levin2006} can be
extended to the R\'enyi entropies for which it has been shown to coincide with the total quantum dimension~\cite{Flammiaetal2009}. Concerning the
relation of the entanglement spectrum to quantum phase transitions, it has been recently shown that a particular subset of such spectrum, that is
the difference between the two largest eigenvalues of the reduced density matrix, the so-called Schmidt gap, plays the role of an effective order
parameter in one-dimensional quantum spin-$1$ models endowed with Haldane topological phases~\cite{DeChiara2012}.

Above all, there is a growing awareness that the entanglement properties of quantum ground states provide the most fundamental characterization
of quantum phases of matter beyond the traditional approach based on symmetry breaking, especially when considering quantum phases of matter
that are established in the absence of symmetry breaking and local order parameters, and are due, e.g., to the presence of hidden topological
order associated to a non-vanishing topological component of the ground-state entanglement entropies~\cite{Balents2012,Kitaev2006,Levin2006}. In
this perspective, the entanglement spectrum and the topological components of the R\'enyi entropies are being actively investigated in various
problems at the cutting edge of condensed matter physics, including Bose-Hubbard spin liquids~\cite{Isakov2011}, frustrated models on nontrivial
lattice geometries~\cite{Vishwanath2011}, non-Abelian fractional Hall systems~\cite{Li2008}, and low-dimensional gapless
models~\cite{ThomaleArovasBernevig2010,Hastings2012}.

When the R\'enyi entropies are computed for blocks $A$ of $\ell$ spins in the ground state of critical models at zero field, it is observed that,
for order $\alpha > 2$, they are characterized by large sub-leading corrections that violate monotonicity in the size of the block ($\ell$),
while no effect appears in the Ising chain at any value of the transverse
field~\cite{Orus,CardyCalabrese2010,CalabreseCampostrinietal2010,HastingsGonzalezKallinMelko2010}. The presence of this oscillatory behavior has
been explained thus far as a peculiar characteristics of critical models, due to parity effects, i.e. depending on whether the number of spins in
a given partition is even or odd.

In the present work we show that the oscillatory scaling is a universal feature of the entanglement spectra and of the R\'enyi entropies in the ground state of critical and noncritical one-dimensional models and that this behavior is related to the existence of a factorizing field, located always below or coinciding with the critical
field ~\cite{GiampaoloAdessoIlluminati2008,GiampaoloAdessoIlluminati2009,GiampaoloAdessoIlluminati2010},
at which the system Hamiltonian admits at least one fully separable ground state.
Using exact analytical methods for integrable models and numerical methods for non-integrable ones, we show that all one-dimensional models
admitting ground-state factorization exhibit two different scaling regimes, separated by the factorizing field $h_f$. In the region
$h\! < \! h_f$ the Schmidt gap and the R\'enyi block entropies exhibit a non monotonic scaling in the size $\ell$ of the block, while
monotonicity is restored for $h\! >\! h_f$. These oscillations are associated to a series of crossovers between the
two entangled eigenvectors of the reduced block density matrix associated to the two largest eigenvalues. Increasing the size of the block the
Schmidt gap closes exponentially in the symmetry broken phase but with a ratio that increases as $h$ decrease for $h_f \! <\! h\! <\! h_c$. On
the contrary, it becomes independent of $h$ as soon as $h < h_f$.
We will relate this general scaling structure in one dimension to a finite-range, entanglement-driven order that is captured by introducing a global order parameter of entanglement.

The paper is organized as follows. In Section II we introduce the class of $XY$ and $XX$ models in transverse field and discuss, using exact analytic methods, the scaling of the R\'enyi block entropies in the ground state of such models, either gapped or gapless, in the thermodynamics limit. In all cases, we show that the factorizing field $h_f$ separates two different scaling regimes in the behavior of the R\'enyi block entropies, either as functions of the transverse field $h$ or of the size of the block $\ell$. In Section III we analyze the behavior of the entanglement spectrum as a function of $\ell$ and we show how the oscillatory behavior is associated to crossovers between the eigenvectors of the block reduced density matrix corresponding to the two largest eigenvalues. As a consequence, we show how the Schmidt gap, i.e. the difference between the two largest eigenvalues, captures most of the significant aspects of the behavior of the full entanglement spectrum. In Section IV we analyze the detailed features of the oscillatory scaling, showing that it is characterized by a set of frequencies that depend only on the normalized transverse field $h/h_f$. As a consequence, a series of ordered entangled structures in terms of quasi-dimers, quasi-trimers, and quasi-polymers is identified for all fields below the factorization point. This form of quantum order is characterized by a global parameter of entanglement defined as the integral, over all block lengths, of the R\'enyi entropy of infinite order. In Section V we extend our investigation to non-integrable models of the $XYZ$ class. At variance with the $XY$, $XX$, and Ising classes, non-integrable models may not always admit ground-state factorization points. Using numerical methods, in particular the density matrix renormalization group (DMRG) algorithm on systems of large but finite size (up to $128$ spins), we discuss analogies and differences between the scaling of the entanglement spectrum in models that admit and models that do not admit factorized ground states.

\section{Scaling of the R\'enyi block entropies in integrable models}

To set the stage let us first consider the class of translationally invariant, one-dimensional $XY$ spin-1/2 Hamiltonians
\begin{equation}\label{Hamiltonian1}
H_{xy}\! =\! \frac{1}{2}\sum_{i} (1+\gamma) \sigma_{i}^x \sigma_{i+1}^x+ (1-\gamma) \sigma_{i}^y \sigma_{i+1}^y - h\sum_{i} \sigma_{i}^z \, ,
\end{equation}
where $\sigma_{i}^\alpha$ $(\alpha \!= \! x,y,z)$ stands for the spin-$1/2$ Pauli operator on site $i$, $h$ is the external transverse field, and
$\gamma$ is the anisotropy, taking values in the interval $[0,1]$, whose extremes correspond, respectively, to the fully isotropic, gapless $XX$
model and to the maximally anisotropic, gapped Ising chain. For every $\gamma$, when $h$ takes the value
$h_f \! =\! \sqrt{1\!-\!\gamma^2}$ the system admits two degenerate, fully factorized ground states that are products of single-site
states~\cite{GiampaoloAdessoIlluminati2008,GiampaoloAdessoIlluminati2009}.
For anisotropy $\gamma \neq 0$ such local states are not, in general, eigenstates of $\sigma^z$. Hence, their tensor product is not an eigenstate of the parity operator along $z$.
On the contrary, their coherent linear symmetric superpositions (with positive or negative relative sign) define two degenerate ground states of definite parity (respectively even or odd) whose entanglement, in general, does not vanish.
For $\gamma \! = \!0$ the two factorized ground states collapse into a single state that preserve all the symmetries of the Hamiltonian.
In the thermodynamic limit, regardless of the value of $\gamma$, at $h_{c}\! =\! 1$ the system undergoes a quantum phase transition
in the $XY$-plane. For $\gamma\!>\!0$ and $h\!<\!h_c$ the system is characterized by a two-fold degenerate ground space and a gapped energy
spectrum with a magnetic order along the $x$ axis, while for $\gamma\!=\!0$ and
$h\!<\!h_c$ the ground state is unique and the energy spectrum is gapless.

The presence of possible degeneracies in the ground space and the fact that different ground states may be characterized by different entanglement properties require, before proceeding further, that we fix the class of ground states on which the investigation shall be focused. Indeed, in order to capture the possible universal aspects in the scaling behavior of the entanglement spectrum and of the R\'enyi block entropies, one needs to treat the gapped case $\gamma \neq 0$ consistently with the isotropic case $\gamma=0$ and with the paramagnetic phase $h > h_c$, since in these two instances the system Hamiltonian admits only a single, non-degenerate ground state that preserves all the symmetries. Therefore, from now on we will consider, also in the magnetically ordered phases of gapped models ($\gamma \neq 0$ and $h < h_c$),
only the class of ground states that preserve all the symmetries of the Hamiltonian, including parity. That is, we will consider only states of definite fixed parity, either even or odd. Of course, also the case of states with no definite parity is physically very relevant. On the other hand, it is technically much more complex, since in such instance one needs to resort to challenging numerical investigations even for the simplest, exactly solvable Hamiltonians like the one-dimensional $XY$ model. Therefore the case of parity non-conserving ground states will be attacked and investigated carefully in a separate work.

Models of the $XY$ and $XX$ class play a relevant role in the field of quantum statistical mechanics because they are exactly solved by resorting to the Jordan-Wigner
transformations of the spin operators into pseudo-fermion
operators~\cite{LiebSchultzMattis,BarouchMcCoyDresded1970,BarouchMcCoy1971,VidalLatorreRicoKitaev2003,LatorreRicoVidal2004}.
Indeed, given a bipartition of our one-dimensional lattice of total size $N$ into a block
of $\ell$ spins and a remainder of size $N - \ell$, it is always possible to obtain,
for ground states of fixed parity, an exact analytical expression of
the reduced block density matrix $\rho_{\ell}$, i.e. of the block reduced state obtained by tracing
the ground state over the degrees of freedom of the remainder. Knowledge of $\rho_{\ell}$ allows to
compute exactly the associated R\'enyi block entropies, defined as:
\begin{equation}
S_{\alpha} (\ell) = \frac{1}{1-\alpha} \ln \left[ \Tr \left( \rho_{\ell}^{\alpha} \right) \right] \; .
\label{Renyi}
\end{equation}

\begin{figure}
\includegraphics[width=8.cm]{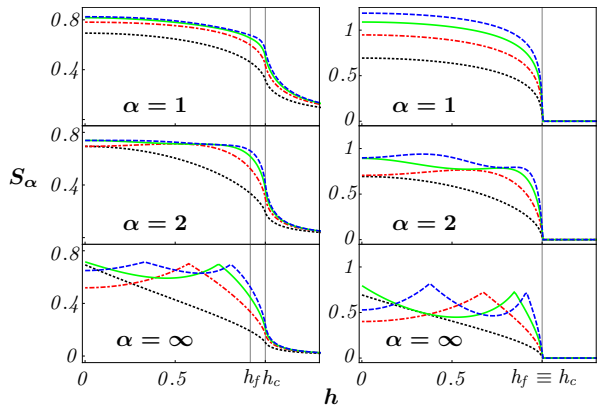}
\caption{R\'enyi block entropies $S_{\alpha}$ as functions of the transverse field $h$ for blocks of different sizes $\ell$, respectively for
gapped ($\gamma\!=\!0.4$ left column) and gapless ($\gamma\!=\!0$ right column) models. Dotted black line: $\ell\!=\!1$; dot-dashed red line:
$\ell\!=\!2$; solid green line: $\ell\!=\!3$; dashed blue line: $\ell\!=\!4$. From top to bottom: plot of the R\'enyi entropies $S_1$, $S_2$, and
$S_{\infty}$.}
\label{Figxyfunh}
\end{figure}

In Fig.~\ref{Figxyfunh} we report the behavior of the R\'enyi entropies $S_\alpha(\ell,h)$, both in the gapless and gapped cases, as functions of the external field $h$ for blocks of spins of different size $\ell$. All calculations are exact and are performed for the one-dimensional $XY$, $XX$, and Ising models in the thermodynamic limit.
We observe that both in the critical and noncritical cases the entropy $S_{1}(\ell,h)$, i.e. the block von Neumann entropy, exhibits a monotonic
behavior both in $h$ and in $\ell$. However, as soon as $\alpha > 1$ increase an oscillatory behavior is observed in the region $h\!<\!h_f$.
Specifically, for $1 \!< \!\alpha \!\leq \! 2$ and $h \! < \! h_f$ the R\'enyi entropies violate monotonicity in $h$ while remaining
monotonically non-decreasing in $\ell$: $S_{\alpha} (\ell \!+\! 1,h)\! > \! S_{\alpha} (\ell,h)$. The amplitude of the oscillations increases as
$\ell$ and $\alpha$ increase. For $\alpha\!>\!2$ the scaling becomes non-monotonic also in the block size:
$S_{\alpha} (\ell \! +\! 1,h)\! >\! S_{\alpha} (\ell,h)$. As already discussed previously, notice that at the factorizing field $h_f$, the entanglement does not vanish, in general, in the parity preserving ground states. However, unambiguous signatures of factorization are readily identified also in the symmetry protected sectors. In particular, for every value of $\ell$, at factorization the block entanglement in parity-symmetric ground states
becomes independent on the relative distance between the different spins in the block. Viceversa, all entanglement-related quantities vanish, correctly, in the non-symmetric fully separable ground states at the factorization point.
\begin{figure}[t]
\includegraphics[width=8.cm]{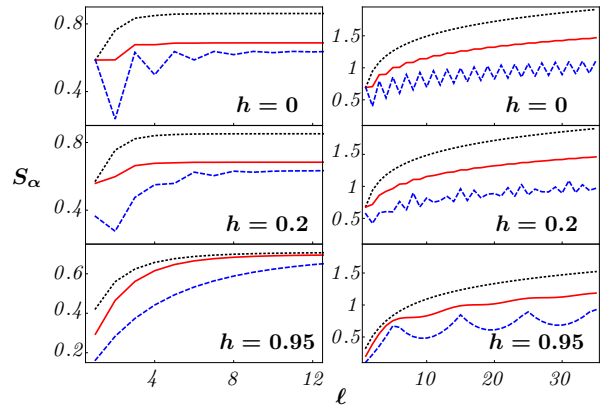}
\caption{R\'enyi block entropies $S_{\alpha}$ as functions of the block size $\ell$ for different values of the index $\alpha$ and of the
external field $h$. Left column: gapped case with $\gamma\!=\!0.4$. Right column: gapless case ($\gamma\! = \! 0$). Black dotted line:
$\alpha\! = \!1$; red solid line:
$\alpha\!=\!2$;  Blue dashed line: $\alpha \!= \!\infty$. Left column, uppermost and central panels: non-monotonic behavior of
$S^\alpha(\ell,h)$ below factorization ($h\!<\!h_f\!\simeq \!0.916$). Lowermost panel: restored monotonic behavior above factorization
(\mbox{$h\! = \!0.95 \!> \!h_f$}), yet below criticality. Panels on the right column: oscillatory scaling in the gapless case, for all values of
$h$ up to factorization that coincides with criticality: $h_f\!\equiv\!h_c$.}
\label{Figxyfunm}
\end{figure}

The central role of the factorizing field for the scaling of the R\'enyi block entropies is illustrated also in Fig.~\ref{Figxyfunm}, where they
are analyzed, for different value of the external field, as functions of the block size $\ell$ both in gapped and gapless models.
For $h \! < \! h_f$ the entropies of order $\alpha \! > \! 2$ exhibit damped oscillations that violate the
area law behavior. As $h$ is increased, the frequency and the amplitude of the oscillations decrease; the area-law monotonic scaling is restored
exactly at factorization. Oscillations are exponentially damped in noncritical (anisotropic) models and only weakly algebraically damped in
critical (isotropic) systems
As the monotonic area-law behavior is restored at the factorization point,
which, for gapped models, means well before the quantum critical point,
the oscillatory scaling behavior must be directly related to the patterns
of entanglement in the ground state, independently of the onset of the long-range
magnetic order at the critical point.

\section{Scaling of the entanglement spectrum}

\begin{figure}[t]
\includegraphics[width=8cm]{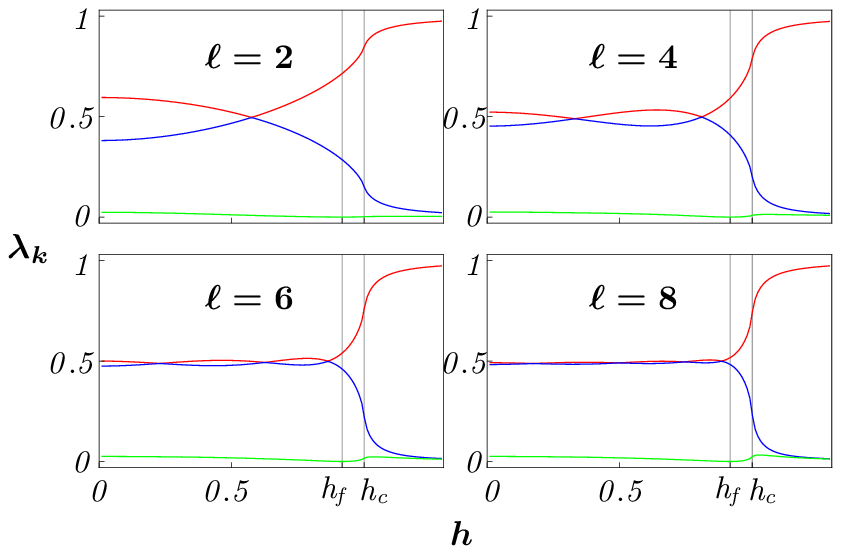}
\includegraphics[width=8cm]{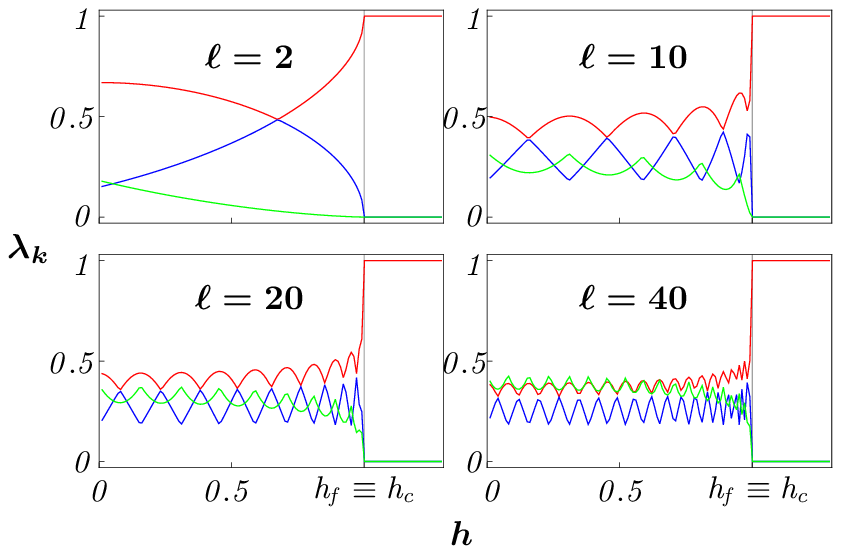}
\caption{Entanglement spectrum $\{ \lambda_k \}$ as a function of the external field $h$ for
different values of the block size $\ell$. Upper panel: anisotropic gapped models ($\gamma\!=\!0.4$).
Lower panel: isotropic gapless models ($\gamma\!=\!0$). In both cases the relevant contributions
are mainly due to the two largest eigenvalues (red line and blue line) compared to the sum of the remaining eigenvalues (green line),
the more so as factorization and criticality are approached. Oscillations increase with increasing size of the block, and terminate
asymptotically at the factorization point. Criticality is signaled by the opening of the Schmidt gap between the two largest eigenvalues.}
\label{EntanglementSpectrum}
\end{figure}

In order to gain a full understanding of the anomalous scaling in relation to the patterns of ground-state entanglement, we turn now to the
analysis of the entanglement spectrum, that is, the entire set $\{ \lambda_k \}$ of the eigenvalues of the reduced density matrix
$\rho_{\ell}$.
The oscillatory behavior of the R\'enyi entropies becomes ever more pronounced as the order $\alpha$
grows. Correspondingly, it follows from the definition of the R\'enyi entropies, Eq.~(\ref{Renyi}), that the
relative weight of the largest eigenvalues of the reduced density matrix $\rho_{\ell}$ increases. Since the characterizations either in terms of
the full entanglement spectrum or in terms of the entire hierarchy of the R\'enyi entropies must be equivalent, we may expect that only the
largest eigenvalues of $\rho_{\ell}$ will contribute significantly to the scaling of the R\'enyi entropies.
Indeed, this is verified both in the gapped and gapless cases, as shown in Fig.~\ref{EntanglementSpectrum}.

From Fig.~\ref{EntanglementSpectrum} we see that the two largest eigenvalues in the entanglement spectrum exhibit an oscillating scaling
behavior analogous to that of the R\'enyi entropies, with growing frequency of the oscillations as the size of the partition increases. Moreover,
for partitions of increasing size $\ell$ the ending point of the oscillations converges, asymptotically, to the
factorization point $h_f$, where $h_f < h_c$ in gapped systems and coincides with it in critical model.
The approach to quantum criticality is signaled by the opening of the Schmidt gap, i.e. the difference between the two largest
eigenvalues~\cite{DeChiara2012}.
Classical saturation to a paramagnetic product ground state at high fields $h$ occurs asymptotically for $h > h_c$ in the gapped case,
and instantaneously at $h_c \equiv h_f$ in the gapless case.

Given these results, besides considering the entire entanglement spectrum, it is sensible to analyze in more detail the scaling of the Schmidt
gap $\Delta$, whose importance has been recently discussed in the analysis of the quantum phase diagram of some one-dimensional spin-1
models~\cite{DeChiara2012}.
\begin{figure}
\includegraphics[width=8cm]{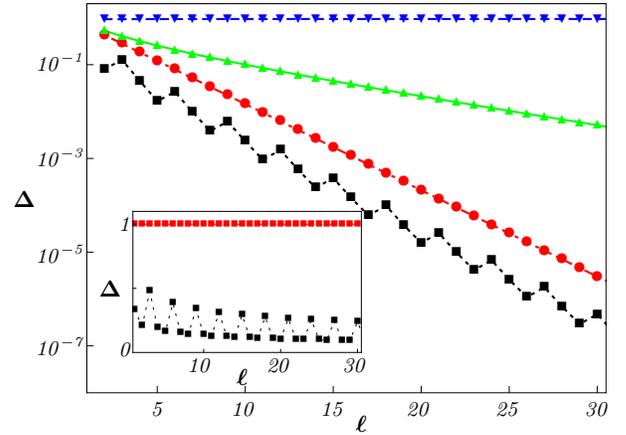}
\caption{Schmidt Gap $\Delta$ as function of the block size $\ell$ for different values of $h$ and $\gamma$. Main panel:
gapped case ($\gamma\!=\!0.4$). Black squares: \mbox{$h\!=\!0.5 h_f \! \simeq \! 0.458$}; red circles: $h\!=\!h_f\!\simeq\!0.916$;
green up-triangles: $h_f\!<\!h\!=\!0.95\!<\!h_c$; blue down-triangles: $h_c\!<\!h\!=\!1.2$. Inset: gapless case ($\gamma\!=\!0$).
Black squares: \mbox{$h\!=\!0.5$}; red circles: \mbox{$h\!=\!h_f\!\equiv \!h_c\!=\!1$}.}
\label{SchmidtGap}
\end{figure}
In Fig.~\ref{SchmidtGap} we report the behavior of the Schmidt gap $\Delta$ as function of the block size $\ell$ for different value of the
external field $h$ both in critical and non-critical models. As $\ell$ increases, the Schmidt gap exhibits an overall envelope that is exponentially decreasing in the ordered phase, $h \! < \! h_c$, in full agreement with the presence of a gap in the energy spectrum.
On the other hand, besides the overall exponential envelope, Fig.~\ref{SchmidtGap} shows two clearly different behaviors: one for
$h_f \! < \! h \! < \!h_c$ and one for $h\! <\! h_c$. In the first region the Schmidt gap undergoes a net exponential decay as $h$ decreases.
When the system enters the region $h \! < \!h_f $ an oscillating behavior is superimposed on the exponential envelope, due to a series of
crossovers between the eigenvectors corresponding to the two largest eigenvalues of the reduced block density matrix. Contrary to what happens
in the first region, the argument of the exponential decay becomes independent of the external field. Being independent of $h$,
the argument of the exponential can then be evaluated analytically at factorization. We obtain that $\Delta(\ell)$, for \mbox{$h\!\le\! h_f$} obeys the relation
$\Delta(\ell) \!\! =\! \! \chi(\gamma)^{-\ell} f(h\!,\!\gamma\!,\!\ell)$, where
$\chi(\gamma) \!\! =\! \! \frac{1}{2 \pi}\! \left|\!  \int_0^{2 \pi} \mathrm{sign} [\cos(\phi) \! - \! (1-\gamma^2)^{1/2} \! + \!
i \gamma \sin(\phi) ] d \phi \right|$, and $f(h\!,\!\gamma\!,\!\ell)$ contains  slowly, non-exponentially decaying oscillating terms and tends to unity as $h$ tends to $h_f$.

\begin{figure}
\includegraphics[width=8cm]{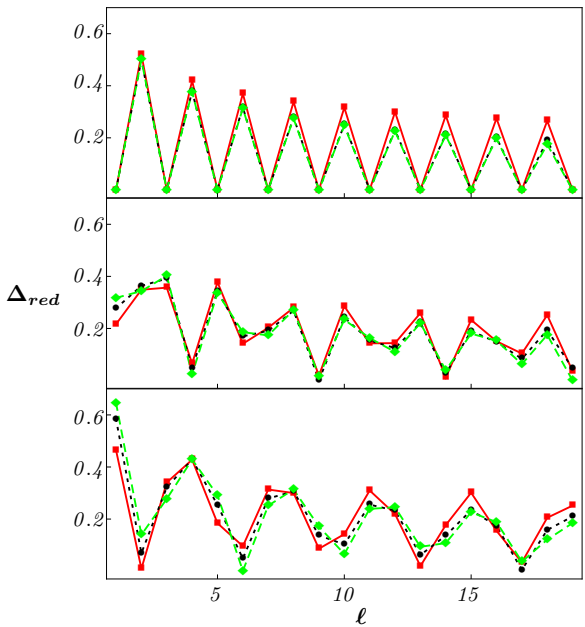}
\caption{Reduced Schmidt gap $\Delta_{red} = \Delta(h,\gamma)/\Delta(h_f,\gamma)$ as a function of the block size $\ell$ for different values of
the external field $h$ and the anisotropy $\gamma$. Uppermost panel: $h=0$. Central panel: $h=h_f/3$. Lowermost panel:$h=(2h_f)/3$. In all
panels: red solid line: gapless case ($\gamma = 0)$; black dotted line: gapped case with $\gamma=0.4$; green dashed line: gapped case with
$\gamma=0.8$. The equal oscillation frequency is a universal effect common to critical and noncritical models.
}
\label{ResidualSchmidtGap}
\end{figure}

In the gapless case, see inset of Fig.~\ref{SchmidtGap}, for all $h<h_c$ the Schmidt gap exhibits a power law decay modulated by an oscillating
dependence on $\ell$, in agreement with the fact that the system is in a critical region and that, being $h_f=h_c$, the external field lies
always below its factorizing value. These observations indicate that there are universal aspects in the scaling of the entanglement spectrum in
one dimension that are independent of the critical or non-critical nature of the systems considered. These universal effects can then be
captured, as reported in Fig.~\ref{ResidualSchmidtGap}, by analyzing the reduced Schmidt gap
$\Delta_{red} \equiv \Delta(h,\gamma)/\Delta(h_f,\gamma)$, i.e. the ratio of the Schmidt gap parametrically dependent on the external field
$h$ and on the anisotropy $\gamma$ to the Schmidt gap evaluated at the factorizing field $h_f$, as a function of the block size $\ell$.
Fig.~\ref{ResidualSchmidtGap} shows that the oscillation frequency in the scaling of the reduced entanglement spectrum is a universal aspect
common to critical and noncritical models, the former differing from the latter only in the overall exponentially decaying envelope whose presence depends on the existence of a gap in the energy spectrum.
The non-perfect superposition of the different curves is partly due to the finite numerical precision and likely also due to the fact that this is an effective, emerging form of universality.

\begin{figure}
\includegraphics[width=8cm]{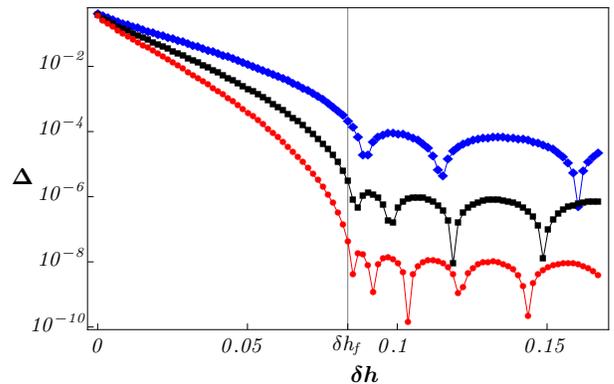}
\caption{Schmidt gap $\Delta$ as a function of $\delta h = |h - h_c|$ for the $XY$ model with anisotropy $\gamma = 0.4$ in the thermodynamic
limit, for different values of the block size $\ell$. Blue diamonds: $\ell = 20$; black squares: $\ell = 30$; red dots: $\ell = 40$.}
\label{SchmidtGapfuncofh}
\end{figure}

The scaling of the Schmidt gap with the external field $h$ is reported in Fig.~\ref{SchmidtGapfuncofh}. It reports the Schmidt gap as a function
of $\delta h = |h - h_c|$ for different values of the block size in the thermodynamic limit. The frequency of the oscillations increases for
growing $\ell$. The oscillatory scaling terminates at the value of $\delta h$ corresponding to the factorizing field, and $\Delta$ opens,
independently of $\ell$, at the critical point $h_c$.

\section{Formation of ordered entangled structures and order parameter of entanglement}

The above results for the scaling of the entanglement spectrum and of the R\'enyi entropies imply that
the factorizing field $h_f$ is itself to be considered a quantum critical point, of a different nature with respect to the standard one
$h_c$ and coinciding with it for gapless models. Exactly at $h \! = \! h_f$ one-dimensional systems undergo a further phase change beyond the
quantum phase transition occurring at the critical point $h_c$. In the region $h \! < \! h_f \! \leq \! h_c$ a finite-range, entanglement-driven
order is established that is superimposed to the global magnetic one driven by spontaneous symmetry breaking.

The implications of the anomalous scaling of the entanglement spectrum for a transitions between different patterns of entanglement in the
ground state can be understood by recalling that the R\'enyi entropy of infinite order \mbox{$S_{\infty} = - \ln{\lambda_{max}}$} coincides with
the single-copy bipartite entanglement~\cite{Eisert2005,Peschel2005} and is monotonic in the bipartite geometric entanglement
\mbox{$\mathcal{E}_G^{(2)} = 1 - \lambda_{max}$}~\cite{Blasoneetal2008}, where $\lambda_{max}$ is the largest eigenvalue of the ground-state
block reduced density matrix $\rho_{\ell}$.
Given a system of total size $N$, $\mathcal{E}_G^{(2)}$ is the minimum distance between the ground state and the set of pure bi-separable
states $\ket{\!\psi_{\ell}} \; \otimes \! \ket{\!\phi_{N - \ell}}$, where $\ket{\!\psi_{\ell}}$ is a state of a block $\ell$, and
$\ket{\!\phi_{N - \ell}}$ is a state of the remainder of the chain. From the last panel of Fig.~\ref{Figxyfunm} one sees that above
factorization $S_{\infty}$ and hence also $\mathcal{E}_G^{(2)}$ increase monotonically in $\ell$.
On the contrary, when $h < h_f$, Fig.~\ref{Figxyfunm} show that $\mathcal{E}_G^{(2)}$ and hence the distance of the ground state from the set of biseparable states
oscillate as a function of $\ell$, with the amplitude of the oscillations increasing for decreasing $h$ and $\gamma$.
In particular, recalling the definition of $\mathcal{E}_G^{(2)} $, we see from Fig.~\ref{EntanglementSpectrum} that at sufficiently small fields the ground state is closest to the bi-separable state
$\ket{\!\psi_{2}} \; \otimes \! \ket{\!\phi_{N - 2}}$.
This implies that well below factorization ($h \ll h_f$) the ground state tends to order in dimerized domains, i.e. entangled structures that involve only two spins in a maximally entangled Bell state. A similar oscillating behavior in the eigenstates of the block reduced density matrices is observed in models that possess exactly dimerized ground states, either due to the presence of competing interactions of different spatial range or due to complex spatial patterns of the nearest-neighbor interaction couplings, as is the case, respectively, of the $J_1 - J_2$ model at the Majumdar-Ghosh point~\cite{MajumdarGhosh} and of models admitting long-distance end-to-end entanglement in the ground state~\cite{LDEcv1,LDEcv2}. However, at variance with such models, in the class of the $XY$ models with nearest-neighbor interactions and spatially constant coupling amplitudes, the oscillatory behavior holds only for the R\'enyi entropies of order $\alpha \geq 2$ and therefore dimerization is not exact. The fact that dimerization is only partial is due to the fact that the relative weight of all the remaining eigenstates of the reduced density matrix with respect to the one of largest amplitude never becomes so small to be put to zero, contrary to what happens, e.g., in the \mbox{Majumdar-Ghosh} model for which it vanishes at each fixed value of $\ell$.

From Fig.~\ref{ResidualSchmidtGap} we see that increasing the strength of the transverse field from values $h \ll h_f$ to values progressively closer to the factorization point, the frequency of the oscillations reduces progressively. This change in the periodicity of the oscillations signals the growth in the spatial dimension of the ordered entangled structures. The latter gradually evolve from strongly entangled domains of quasi-dimers, leaving place to larger structures ranging from quasi-trimers up to quasi-polymers, until their size diverges exactly at the factoring field. In other words, as one moves away from vanishingly small values of the external field and approaches the factorization point, the associated entangled structures become spatially more extended. At the same time, as these extended structures develop close to factorization and involve ever larger number of sites, the associated entanglement becomes smaller and smaller in each pair of sites, in full agreement with the behavior of the pairwise concurrence (entanglement of formation) close to a factorization point~\cite{Amico2006}.

\begin{figure}
\includegraphics[width=7.cm]{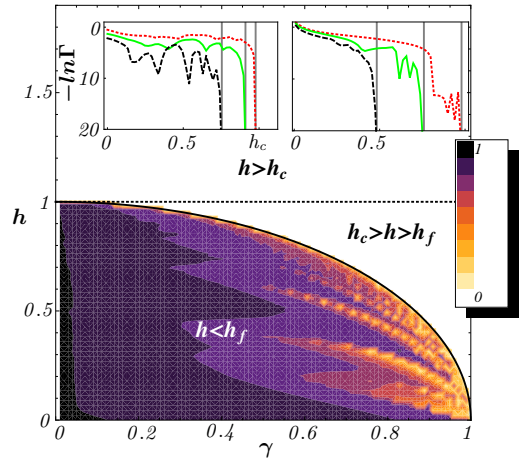}
\caption{Order parameter of entanglement $\Gamma$ as a function of $h$ and $\gamma$. Left inset: one-dimensional projection as a function of
$h$ for different values of $\gamma$. Red dotted line: $\gamma=0.2$; green solid line: $\gamma = 0.4$; black dashed line: $\gamma = 0.65$.
Right inset: one-dimensional projection as a a function of $\gamma$ for different values of $h$. Red dotted line: $h=0.2$; green solid line:
$h = 0.65$; black dashed line: $h = 0.87$. In both insets the vertical grid lines correspond to the factorization points.}
\label{OPFig}
\end{figure}

As the oscillatory behavior is maximized by the single-copy entanglement $S_{\infty}(\ell,h)$, the finite-range, 
entanglement-driven order below factorization is naturally characterized by an order parameter of entanglement defined as the integral of
$S_{\infty}(\ell,h)$ over partitions of all sizes:

\begin{equation}
\Gamma = - \sum_{\ell=1}^{\infty}
\min [ S_{\infty}(\ell + 1,h) - S_{\infty}(\ell,h), 0 ] \; .
\label{OP}
\end{equation}
In Fig.~\ref{OPFig} we report the contour plot of $\Gamma$ for the class of one-dimensional $XY$ models. It vanishes identically for $h > h_f$.
Approaching the Ising limit it is strongly reduced, corresponding to strongly damped oscillations. Approaching the $XX$ limit, as the
oscillations become long-ranged, $\Gamma$ increases and diverges exactly at $\gamma = 0$, consistently with the fact that in gapless models below
the critical point the oscillations persist, with weak algebraic decay, also for large $\ell$.
Eq.~(\ref{OP}) and Fig.~\ref{OPFig} show that the quasi-dimerized order below factorization is exclusively due to the entanglement properties of
each individual ground state and therefore, unlike magnetic order, it is not a consequence of ground-state degeneracy and symmetry breaking.


\section{Scaling of the entanglement spectrum in non-integrable models}

We will now show that the results of the above analysis extend as well to generic non-integrable models of the $XYZ$ and Heisenberg type:
\begin{equation}\label{Hamiltonian2}
H_{xyz} = \frac{1}{2}\sum_{i,l} J_x \sigma_{i}^x \sigma_{l}^x + J_y \sigma_{i}^y \sigma_{l}^y + J_z \sigma_{i}^z \sigma_{l}^z - h\sum_{i}
\sigma_{i}^z \, .
\end{equation}
Here $J_\mu$ are the spin-spin couplings along the $\mu\!=\! x,y,z$ directions and, without loss of generality, we set $J_x\!=\!1\ge\!
|J_y|,|J_z|$. These systems undergo a quantum phase transition at $h \!= \!h_c$, while, contrary to the integrable cases discussed above,
factorized ground states do not necessarily exist. At each finite value of the external field, the factorizing field is defined as
$h \! = \! h_f \! =\! \sqrt{(J_x\!+\!J_z)(J_y\!+\! J_z)}$, and therefore it exists if and only if
$J_z \! \ge \! -J_y$~\cite{GiampaoloAdessoIlluminati2008,GiampaoloAdessoIlluminati2009}. The models in Eq.~(\ref{Hamiltonian2}) are not exactly
solvable. In order to determine the block reduced density matrices we diagonalized the system by means of the Density Matrix Renormalization
Group (DMRG)~\cite{ReviewSchollwoeck,JCP,White} applied to open chains of up to $128$ spins. We kept up to $m=16$ states of the reduced density
matrix, letting the truncation error stay well below $10^{-7}$ at each step.

For models admitting a factorization point, the results of the $XY$ case carry over essentially unmodified to the general $XYZ$ instance.
\begin{figure}
\includegraphics[width=7.cm]{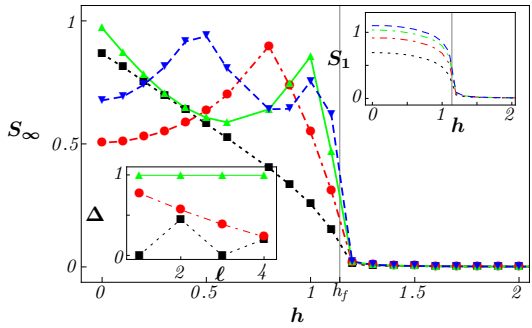}
\caption{R\'enyi entropies and entanglement spectrum in non-integrable one-dimensional $XYZ$ models with $J_x\!=\!1$,
$J_y\!=\!0.7$, $J_z\!=\!0.3$, and $h_f\! \simeq\! 1.14$. Main plot: single-copy entanglement $S_{\infty}$ as a function of $h$ for different
values of the block size $\ell$. Black Squares: $\ell\! =\! 1$; red circles: $\ell\! = \!2$; green up-triangles: $\ell\! = \!3$;
blue down-triangles: $\ell \!=\! 4$.Uppermost right inset: von Neumann entropy $S_1$ as a function of $h$ for different values of $\ell$.
Lowermost left inset: Schmidt gap $\Delta$ as a function of $\ell$ at different values of $h$. Black squares $h\!=\!0$; red circle
$h\!=\!1.145>h_f$; green up-triangles $h=2>h_c$.}
\label{Figwith}
\end{figure}
In Fig.~\ref{Figwith} we report the scaling of the R\'enyi entropies in $XYZ$ models that possess a factorizing field
$h_f$ ($J_z \! \ge \! -J_y$) and the scaling of the Schmidt gap as a function of the block size $\ell$ for different value of the external field.
The qualitative behavior of these two quantities is analogous to that of the $XY$ models and confirms the existence of an
entanglement-induced order of quasi-dimerized domains in the region $h\! < \! h_f$.
\begin{figure}[t]
\includegraphics[width=7.cm]{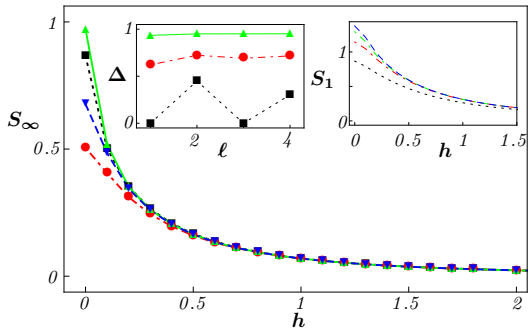}
\caption{R\'enyi entropies and Schmidt gap in $XYZ$ models with $J_x\!=\!1$, $J_y\!=\!-0.7$, $J_z\!=\!-0.3$, that thus do not admit
factorization points. Main plot: Single-copy entanglement $S_{\infty}$ as a function of the external field $h$ for different values of the block
size $\ell$. Black Squares: $\ell\! =\! 1$;
red circles: $\ell\! = \! 2$; green up-triangles: $\ell \!=\! 3$; blue down-triangles: $\ell\! =\! 4$.
Right inset: von Neumann entropy $S_1$ as a function of $h$ for different values of $\ell$.
Left inset: Schmidt gap $\Delta$ as a function of $\ell$ at different values of $h$. Black squares: $h\!=\!0$; red circles: $h\!=\!1.14$;
green up-triangles: $h=2$.}
\label{Figwithout}
\end{figure}

For models that do not admit a factorized ground state at any finite value of the magnetic field, the scaling behavior of the R\'enyi entropies and of the entanglement spectrum is rather different, as reported in  Fig.~\ref{Figwithout}. At variance with models that do admit
ground-state factorization, the area-law scaling with the size of the block, $S_{\alpha} (\ell \!+ \!1,h) > S_{\alpha} (\ell,h)$, appears to be
\emph{always} violated for \emph{all} values of $h$, as one can see from Fig.~\ref{Figwithout}. However, all the R\'enyi entropies behave smoothly and do not acquire local maxima. On the other hand, when considered as functions of the block size $\ell$ (at fixed external field $h$),
the block entropies exhibit the same oscillating behavior as in models with factorized ground states. These features can be intuitively understood by considering that in these models factorization, so to speak, is shifted away towards infinitely large values of $h$, indeed corresponding to a classical saturation. The latter is the physical limiting situation in which an extremely strong transverse field orients all spins along the field direction, resulting in a classical paramagnetic state that is naturally a product of single-site paramagnets, with all quantum correlations washed away. Indeed, as we can see from the left inset of Fig.~\ref{Figwithout}, the true quantum oscillations in the Schmidt gap appear to be confined up to a value of the external field $\tilde{h}\!=\!\sqrt{\!(J_x\!+\!J_z)(\!-J_y\!-\! J_z)}$, which, somehow, corresponds to an \emph{inverted, de-complexified} factorization point. Therefore, a sort of would-be factorizing field seems to play a very important role also in models that do not admit a physical factorized ground state.

A complete, rigorous confirmation of these behaviors for all values of the transverse field $h$ in non-integrable models is beyond our current numerical possibilities for two reasons: firstly, the
differences between R\'enyi entropies of the same order $\alpha$ and different partition size $\ell$ fall off extremely rapidly with $h$; moreover, the limitations on current algorithms do not allow a reliable analysis for partitions of size $\ell\!>\!4$. This is due to the fact that the algorithm is based on the direct computation of the two-site density matrix of the two central sites of the DMRG ansatz. The further density matrices, e.g. the four-site density matrix, are obtained by repeated merging of two physical sites in one computational site. This protocol is very effective in obtaining all quantities of interest but, as the local dimension of the computational sites increases exponentially with the number of physical sites grouped together, it prevents at the moment going beyond blocks of size $\ell\!>\!4$. If further confirmed by future more powerful numerical analysis, the reported behavior would imply that the entanglement-induced order can exist also in disordered phases with vanishing local order parameters.

\section{Discussion and outlook}

We have investigated the scaling behavior of the entanglement spectrum and of the R\'enyi entanglement entropies in the ground state of
one-dimensional gapped and gapless quantum spin models. We have showed that a violation of the area law scaling behavior occurs both in the
critical and non-critical cases for all R\'enyi entropies of order $\alpha \!>\! 2$ and for external fields $h \!<\! h_f \!<\! h_c$ in models
admitting ground-state factorization. An analogous behavior is observed in the Schmidt gap, that is the difference between the two largest
eigenvalues of the block reduced density matrix.

The existence of a factorizing field and the associated anomalous scaling of the entanglement spectrum correspond to the existence of a quantum
regime superimposed to and distinct from the ordered and disordered phases associated to the existence of ground states breaking some symmetry of
the Hamiltonian. Unlike long-range magnetic order due to spontaneous symmetry breaking, this finite-range, entanglement-driven order does not
rely on the onset of ground-state degeneracy, but is rather due to the entanglement properties of each ground state of fixed parity. The
factorizing field plays a key role in the understanding of the anomalous scaling of the entanglement spectrum by clarifying the origin of this
phenomenon even in the gapless case in which the factorization point and the critical point coincide. It also explains why the anomalous scaling
is not observed in systems, like the Ising model, for which $h_f \!= \!0$. For non-integrable models that do not admit factorization, the oscillatory scaling appears to extend to all values of the external field, while the non-monotonic behavior of the Schmidt gap appears to be confined in the region
$h\!<\! \tilde{h}\!=\!\sqrt{(J_x\!+\!J_z)(\!-J_y\!-\! J_z)}$. However, further numerical analyses will be required in order to reach a definite assessment for such models.

As already mentioned in the introduction~\cite{Levin2006,Balents2012}, a new quantum revolution is taking place in condensed matter physics inasmuch as it is being recognized that there are kinds of quantum phases that are truly and uniquely characterized by their entanglement properties, the paramount example being
the quantum spin liquid phase associated to a nonvanishing topological entanglement entropy for some classes of two-dimensional, geometrically frustrated quantum spin systems (e.g. the Heisenberg model on the Kagom\'e lattice). In fact, this fundamental realization has motivated the characterization and classification of quantum phases according to equivalence classes of entanglement: for instance, ground states endowed with topological order are characterized by a constant, topological entanglement entropy of infinite spatial range. Therefore, they cannot be adiabatically connected to ground states with short-range entanglement patterns without crossing a quantum phase transition. In particular, phases endowed with topological
order are not compatible with factorized ground states, the latter being a particular element of the equivalence class of states with short-range entanglement~\cite{Wenlocalunitary,Hammalocalcharacterization}.

The above holds true (although a rigorous proof is still lacking) for two- and higher-dimensional systems, where topological effects con occur independently of the presence or absence of symmetries. In one-dimensional systems, it has been conjectured that topological phases can occur only if the associated ground states share some given symmetry and cannot be adiabatically deformed into each other without crossing a quantum phase transition, if the deformation preserves that symmetry~\cite{Verstraete,Gu,Pollmann}, leading to a classification of gapped symmetric phases in one-dimensional spin systems~\cite{Chen}. This is certainly not the case for the exchange quantum spin-1/2 chains that we have discussed in the present work. On the other hand, the conjecture puts the role of the factorized ground state at the center stage of the investigation, and for models that are thought of possessing true symmetry-protected topological phases in one dimension, our methods for the study of ground-state factorization could be applied in order to verify the conjecture in explicit model Hamiltonians in one dimension.

Frustration, a crucial ingredient for the realization of topological phases in two dimensions, appears to be an important factor in one-dimensional systems as well. We have noticed that the single-copy entanglement $S_{\infty}$ is monotonic in the bipartite geometric entanglement $\mathcal{E}_G^{(2)}$, which in turn has been shown to be a universal lower bound to ground-state frustration, independently of the
spatial dimension~\cite{Giampaolo2011,Marzolino}. This correspondence suggests the existence of an intimate relation between the scaling of the entanglement spectrum, the tendency to form ordered entangled structures below the factorization point, and the frustration of purely quantum origin. In such a perspective, the present investigation might be further fruitfully extended to geometrically frustrated models and systems with topological order and other classes of exotic quantum phases.

Finally, the fact that the factorization point has been identified to constitute the boundary point for different types of scaling of the ground-state entanglement suggests that the presence of factorization might impose fundamental limits to Trotterization and other classes of adiabatic methods~\cite{Lloyd,Wunderlich} for the simulatability of quantum many-body Hamiltonians. This crucial point of great practical
importance will be the subject of forthcoming work.

\section*{Acknowledgements}

One of us (FI) thanks A.~Hamma for inspiring exchanges. One of us (SM) thanks M.~Rizzi for discussions. SDS, SMG, and FI acknowledge support from the EU STREP Project iQIT, Grant Agreement No. 270843. SMG also gratefully acknowledges the Austrian Science Fund (FWF-P23627-N16). SM acknowledges the PwP project (www.dmrg.it) for the DMRG code, the DFG (SFB/TRR~21), the bwGRiD, and the EU-SIQS
for support.

\end{document}